# Supplementary Feedforward Voltage Control in a Reconfigurable Distribution Network

Young-Jin Kim

***Abstract*—Network reconfiguration (NR) has attracted much attention due to its ability to convert conventional distribution networks (DNs) into self-healing grids. This paper proposes a new strategy for real-time voltage regulation (VR) in a reconfigurable DN, whereby optimal feedforward control of synchronous and inverter-based distributed generators (DGs) is achieved in coordination with the operation of feeder line switches (SWs). This enables preemptive compensation of upcoming deviations in DN voltages caused by NR-aided load restoration. A robust optimization problem is formulated using a dynamic analytical model of NR to design the feedforward voltage controllers (FVCs) that minimize voltage deviations with respect to the $H_\infty$ norm. Errors in the estimates of DG parameters and load demands are reflected in the design of optimal FVCs through polytopic uncertainty modeling, further improving the robustness of the proposed VR strategy. Small-signal analysis and case studies are conducted, demonstrating the effectiveness of the optimal robust FVCs in improving real-time VR when NR is activated for load restoration. The performances of the proposed FVCs are also verified under various operating conditions of a reconfigurable DN, characterized principally by SW operations, network parameter errors, and communication time delays.**

***Index Terms*—Load restoration, network reconfiguration, polytopic uncertainty, robust optimization, voltage control.**

## NOMENCLATURE

***Sets***

| | |
|---|---|
| $d$, $q$ | subscripts for $d$- and $q$-axis variables |
| $i$, $k$, $n$, $v$ | indices for SGs, IGs, buses, and vertices |
| $t$, $T$ | index and total number of sampling time steps |
| $G$, $L$, $N$, $V$ | total numbers of SGs, IGs, buses, and vertices |
| $\mathcal{P}$ | convex polytope set |
| $\|\|\bullet\|\|_\infty$, $\|\|\bullet\|\|_2$ | infinity- and two-norm values of • |
| $\bar{\bullet}$, $\underline{\bullet}$ | maximum and minimum estimates of • |
| $diag(\bullet)$ | block diagonal matrix composed of • |
| $tr(\bullet)$ | sum of the diagonal elements of • |
| $Co(\bullet)$ | convex hull for the set of vertices of • |
| $\sigma(\bullet)$ | singular values of • |

***Matrices, vectors, and scalars***

| | |
|---|---|
| $u(t)$ | NR-initiating signal |
| $H_i(s)$, $M_k(s)$ | transfer functions of the FVCs for SG unit $i$ and IG unit $k$ |
| $K_A$, $L_f$ | exciter amplifier gain and line filter inductance |
| $S_L$, $S_r$ | total load demand and the amount of load to be restored |
| $T_d$ | communication time delay |
| $U_{SGi}$, $U_{IGk}$ | output signals of the FVCs for SG unit $i$ and IG unit $k$ |
| $V_{SGi}$, $V_{IGk}$ | terminal voltage magnitudes of SG unit $i$ and IG unit $k$ |
| $\Delta T_{set}$ | settling time of voltage deviation |
| $\Delta V_{rms}$, $\Delta V_{pk}$ | rms and peak-to-peak voltage deviations |
| $\gamma$ | upper bound of the energy of FVC output signals |
| $\mathbf{d}(s)$ | padé approximation of the time-delay transfer function |
| $\mathbf{A_X}$, $\mathbf{B_V}$, $\mathbf{C_X}$, $\mathbf{D_V}$, $\mathbf{D_L}$ | modeling coefficients of DGs and voltage-dependent loads |
| $\mathbf{A_{DN}}$, $\mathbf{B_{NR}}$, $\mathbf{B_{DG}}$, $\mathbf{C_{DG}}$ | modeling coefficients of a reconfigurable network |
| $\mathbf{A_{FF}}$, $\mathbf{B_{FF}}$, $\mathbf{C_{FF}}$, $\mathbf{U_{FF}}$ | control parameters and output signals of FVCs |
| $\mathbf{A_{OD}}$, $\mathbf{B_{OD}}$, $\mathbf{C_{OD}}$ | coefficients for the overall dynamics of a reconfigurable network including SGs and IGs with optimal FVCs |
| $\mathbf{G}(s)$, $\mathbf{G_d}(s)$ | dynamic responses of $\mathbf{V_{DG}}$ to NR without and with consideration of communication time delays |
| $\mathbf{G_{FF}}(s)$ | dynamic response of FVCs to an NR-initiating signal |
| $\mathbf{I_0}$, $\mathbf{V_0}$ | $dq$-axis currents and voltages in the steady state |
| $\mathbf{I_{DG}}$, $\mathbf{I_L}$ | injection currents of DGs and voltage-dependent loads |
| $\mathbf{\Delta I_T}$, $\mathbf{\Delta Y}$ | variations in injection currents and the admittance matrix |
| $\mathbf{V_{DG}}$ | terminal voltage magnitudes of SGs and IGs |
| $\mathbf{X_{DN}}$, $\mathbf{X_{FF}}$, $\mathbf{X_{OD}}$, | states of a reconfigurable network, optimal FVCs, and their overall dynamics |
| $\mathbf{X_{SG\mathit{i}}}$, $\mathbf{X_{IG\mathit{k}}}$ | states of SG unit $i$ and IG unit $k$ |
| $\mathbf{Y_B}$, $\mathbf{Y_A}$ | admittance matrices before and after NR |
| $\mathcal{J}$, $\mathcal{C}_{1\text{–}3}$, $\mathcal{C}_N$ | objective function and constraints of an optimization problem to design optimal FVCs |
| $\mathbf{Q}$, $\mathcal{R}$, $\mathcal{N}$, $\mathcal{U}$, $\mathcal{V}$, $\mathcal{L}_{1\text{–}5}$ | positive definite matrix for the Lyapunov condition and auxiliary variables for LMI constraints |
| $\tilde{\mathcal{T}}$, $\mathcal{T}$ | congruence transformation matrices |

## I. INTRODUCTION

EXTREME weather events, such as floods and storms, are increasingly threatening the reliability of distribution power grids. In the United States, the costs of weather-related power outages were estimated to range between approximately $25 billion and $70 billion per year during the period from 2003 to 2012 [1]. Over this period, the annual number of major weather-related outages, which affected at least 50,000 customers, increased from less than 40 to more than 80 [2]. Moreover, 90% of outages occurred at the distribution levels [1]–[3]. This emphasizes that improving the resilience of distribution networks (DNs) is of key importance when establishing future smart grids [4], [5].

Dynamic network reconfiguration (NR) has attracted much attention. This enhances the resilience by enabling self-healing operations of DNs. NR changes the topological structure of the DN through on-off operations of line switches (SWs). Distribution line faults can then be isolated, and de-energized loads are re-connected to distribution feeders where load services are sustained by power supplied from main substations and distributed generators (DGs).

Y. Kim is with the Department of Electrical Engineering, Pohang University of Science and Technology, Pohang, Gyungbuk 790-784, Korea (e-mail: powersys@postech.ac.kr).

In most prior studies regarding NR (e.g., [6]–[9]), the operational schedules of SWs were determined in advance, for example, to maximize the restored, critical load demand, while minimizing the time required for load service restoration. However, scheduling was achieved while considering only the steady-state operations of reconfigurable DNs. Only time-invariant or hourly-sampled profiles of load demand were taken into account; DGs were regarded as point sources and their dynamic responses were thus neglected. Consecutive SW operations are very likely to cause sudden variations in load demand, which in turn triggers abrupt fluctuations of DN voltages in the transient state. Given the small capacities and low inertias of DGs, voltage fluctuations can cause further unexpected tripping of DGs and cascading collapses of DN voltages. This implies that it is essential to accurately reflect the dynamic response characteristics of DGs, loads, and bus voltages into NR-aided load restoration.

In [10]–[12], the dynamic responses of DG and load units were reflected into the optimization problems to schedule the optimal operations of SWs, while maintaining DN voltages within acceptable ranges for each SW operation. Synchronous generators (SGs) were mainly considered as the DG units, although their dynamics were somewhat simplified. In [13] and [14], optimal NR scheduling was performed with consideration of the transient variations in DN voltages during load restoration via trial-and-error approaches. The sizes and locations of de-energized loads that can be restored without violating constraints on transient voltage variations were pre-selected through iterative simulation. However, load services were recovered using SGs alone, rather than SGs in cooperation with inverter-based generators (IGs). Moreover, in [10]–[14], real-time DG control was achieved by conventional feedback control loops [15] that came into effect only after DN voltages had substantially deviated because of NR. Thus, current real-time voltage regulation (VR) in a reconfigurable DN can be further improved.

Only a few recent works (e.g., [16]–[18]) have investigated the coordination of DGs and SWs to improve real-time VR during load service restoration. In particular, supplementary feedback control loops were established between SGs and SWs [16] and between IGs and SWs [17], [18]. These allowed adjustment of the terminal voltages of SGs and IGs, for example, by reference to the on-off status and terminal voltages of SWs and the currents flowing through the SWs. The adjustments maintained the differences between the terminal voltages of each SW at zero prior to NR; otherwise, large inrush currents were likely to occur, leading to severe voltage fluctuations. However, such supplementary control is possible only when the feeders of both terminals of the SW are energized. Thus, the method is not applicable to NR-aided load restoration, because the voltages become zero at SW terminals that are connected to interrupted loads. Consequently, the terminal voltages of SGs and IGs still need to be regulated through conventional feedback control, as in [10]–[15].

These issues have motivated the development of new strategies to preemptively regulate DN voltage deviations caused by NR, because NR is commonly performed in a controlled manner. To develop such VR strategies, the literature research gap between the dynamic NR models and their application to DG control needs to be filled first. In [10]–[14], NR was simply modeled as the amount of load to be restored or shed, rather than a change in the network topology itself. This compromises the estimation accuracy of the dynamic responses of DGs and loads to the SW operations involved in NR-aided load restoration. In addition, the uncertainties in estimates of DG and load parameters were not explicitly considered in [6]–[18], although uncertainties are inevitable in real DNs. When uncertainties are neglected, pre-emptive regulation of DN voltages can become practically ineffective, failing to ensure stable operation of DNs.

In this paper, we propose a new strategy for real-time VR of a reconfigurable DN. Optimal feedforward control of SGs and IGs is achieved in coordination with the SW operations to preemptively mitigate voltage deviations at DG terminal buses attributable to NR. The dynamic responses of bus voltages to NR are estimated using an analytical model of a reconfigurable DN; these responses are integrated into a robust optimization problem to design optimal feedforward voltage controllers (FVCs) that minimize voltage deviations. Uncertainties in the estimates of DG parameters and load demands are considered during optimization, further improving the robustness of optimal FVC operations. The FVCs are incorporated in parallel with existing feedback control loops to eliminate steady-state variations in DG terminal voltages. A small-signal analysis and case studies are conducted to assess the performance of the proposed VR strategy using comprehensive models of DGs.

The main contributions of this paper are summarized below:

- To the best of our knowledge, this is the first study to report feedforward control of SGs and IGs in coordination with SW operations to improve real-time VR in a reconfigurable DN during load service restoration through NR.
- A convex optimization problem is formulated to develop the optimal robust FVCs that minimize the upcoming variations in DG terminal voltages due to NR in the sense of $H_\infty$ norm.
- Errors in the estimates of DG parameters and load demands are reflected in the optimization problem using a polytopic uncertainty model, further enhancing the effectiveness and robustness of the optimal FVCs when applied to real DNs.
- Comparative small-signal analysis and numerical case studies are comprehensively conducted under various grid conditions, characterized by SW operations, uncertainty levels, and communication delays.

## II. Fundamentals and Framework

In distribution feeders, NR is achieved through on-off operations of SWs such as sectionalizing switches (SSWs) and tie switches (TSWs). SSWs are installed along individual feeders, and TSWs are installed between feeders. The current practices and standards [19], [20] state that a distribution system operator (DSO) should send either automatically or manually binary signals (zero to one or vice versa) to SSWs and TSWs via communication links when changing on-off status. Note that the switching sequences or schedules are often pre-determined, as discussed in Section I, and the binary signals can serve as NR-initiating signals in the proposed VR strategy. Moreover, in general, DGs regulate their terminal voltages to the reference values in real time, while supplying active and reactive power to distribution feeders. Real-time VR at DG terminal buses facilitates the DSO to support bus voltages across a DN. Conventionally, VR has been achieved using the feedback control loops of individual DGs, commonly by employing proportional-integral (PI) controllers [15], [21].

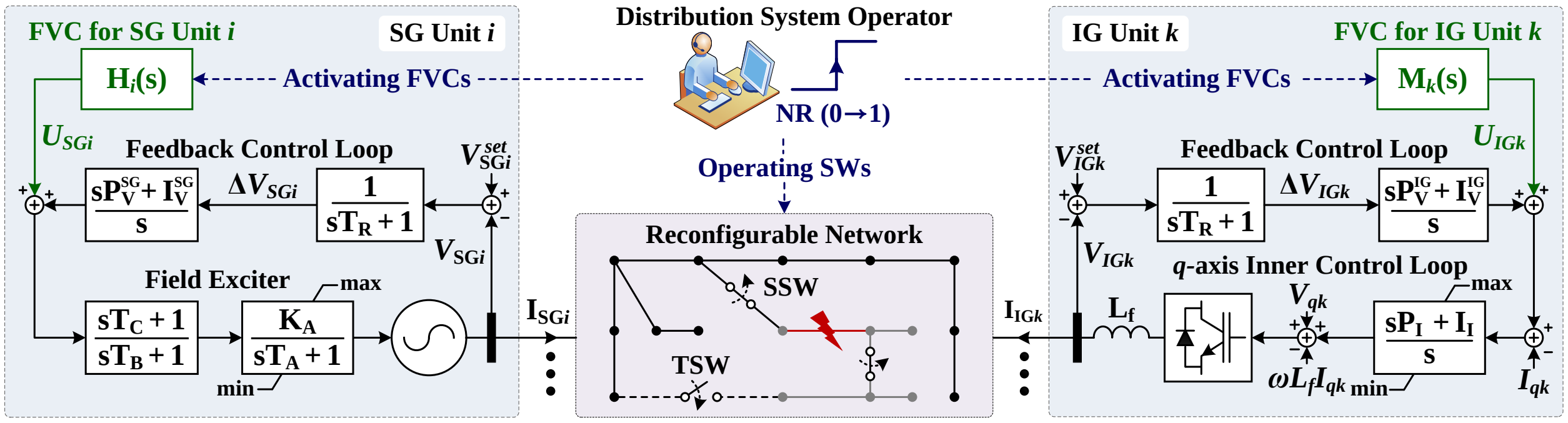

Fig. 1. Schematic diagram of the proposed strategy for real-time VR in a reconfigurable DN including SGs and IGs.

Fig. 1 shows a schematic diagram of the proposed strategy for a reconfigurable DN, wherein the FVCs of SGs and IGs generate reference signals for the field exciters and the $q$-axis inner control loops, respectively, in response to binary signals (or, alternatively, the NR-initiating signals). In Fig. 1, $H_i(s)$ and $M_k(s)$ represent the FVCs for SG unit $i$ and IG unit $k$, respectively. The FVCs are implemented in the same locations as the DGs, and then incorporated in parallel to existing feedback control loops. The FVCs enable SGs and IGs to respond faster and preemptively to forthcoming variations in the terminal voltages caused by NR, thus allowing existing feedback loops to better compensate for remaining voltage variations. This improves the transient stability of bus voltages throughout the DN in a more robust manner than feedback control loops alone.

In this paper, the FVCs are optimized using only information that is commonly available on a reconfigurable DN, SGs, and IGs. The optimal FVC parameters are updated online based on the current load demand and the locations of target SWs to better reflect the time-varying dynamics of a DN, as in the common multi-controller architecture [22]. Delivery of an NR-initiating signal and updating of the FVC parameters are performed only when SW operations are involved. This mitigates the requirement for high-bandwidth communication networks, thus facilitating implementation of the proposed strategy in real DNs.

## III. Design of Optimal Robust FVCs

### A. Dynamic Responses of DG Terminal Voltages to NR

To design the proposed FVCs, the dynamic responses of DG terminal voltages to SW operations are first estimated using an analytical model of the reconfigurable DN. In a previous work [23], the analytical model was developed, wherein NR was considered to be a change in the DN topology itself. This improved the estimation accuracies of network voltage responses, compared with estimates of conventional models where NR was simply regarded as the load demand to be restored or shed. In this paper, the analytical model is further adapted for its application to supplementary feedforward control of SGs and IGs in response to NR-initiating signals. Briefly, (1)–(5) show the analytical model of the reconfigurable DN, given the parameters of the SGs, IGs, voltage-dependent loads, and three-phase distribution lines, as:

$$\Delta\dot{\mathbf{X}}_{\mathbf{DN}}(t) = \mathbf{A}_{\mathbf{DN}}\cdot\Delta\mathbf{X}_{\mathbf{DN}}(t) + \mathbf{B}_{\mathbf{DG}}\cdot\Delta\mathbf{U}_{\mathbf{FF}}(t) + \mathbf{B}_{\mathbf{NR}}\cdot u(t), \quad (1)$$

$$\Delta\mathbf{V}_{\mathbf{DG}}(t) = \mathbf{C}_{\mathbf{DG}}\cdot\Delta\mathbf{X}_{\mathbf{DN}}(t), \quad (2)$$

$$\text{where } \Delta\mathbf{X}_{\mathbf{DN}} = [\Delta\mathbf{X}_{\mathrm{SG}1}, \ldots, \Delta\mathbf{X}_{\mathrm{SG}G}, \Delta\mathbf{X}_{\mathrm{IG}1}, \ldots, \Delta\mathbf{X}_{\mathrm{IG}L}]^T, \quad (3)$$

$$\Delta\mathbf{U}_{\mathbf{FF}} = [\Delta U_{SG1}, \ldots, \Delta U_{SGG}, \Delta U_{IG1}, \ldots, \Delta U_{IGL}]^T, \quad (4)$$

$$\Delta\mathbf{V}_{\mathbf{DG}} = [\Delta V_{SG1}, \ldots, \Delta V_{SGG}, \Delta V_{IG1}, \ldots, \Delta V_{IGL}]^T. \quad (5)$$

Please refer to Appendix A for details. In (1), $u(t)$ is the NR-initiating signal. This can also represent a signal generated at any arbitrary time $t$ without loss of generality, when $\mathbf{A}_{\mathbf{DN}}$, $\mathbf{B}_{\mathbf{DG}}$, $\mathbf{B}_{\mathbf{NR}}$, and $\mathbf{C}_{\mathbf{DG}}$ are accordingly updated prior to NR. Thus, (1)–(5) can be applied to the consecutive operations of SWs. Note that the updating is performed based on the load demand and the SW locations, as discussed in Section II.

### B. Formulation of the Robust Optimization Problem

The proposed FVCs are designed in the form as:

$$\Delta\dot{\mathbf{X}}_{\mathbf{FF}}(t) = \mathbf{A}_{\mathbf{FF}}\cdot\Delta\mathbf{X}_{\mathbf{FF}}(t) + \mathbf{B}_{\mathbf{FF}}\cdot u(t), \quad (6)$$

$$\Delta\mathbf{U}_{\mathbf{FF}}(t) = \mathbf{C}_{\mathbf{FF}}\cdot\Delta\mathbf{X}_{\mathbf{FF}}(t), \quad (7)$$

where $\mathbf{A}_{\mathbf{FF}}$, $\mathbf{B}_{\mathbf{FF}}$, and $\mathbf{C}_{\mathbf{FF}}$ are the control parameters to be optimally determined. The size of $\Delta\mathbf{X}_{\mathbf{FF}}$ is set to be the same as that of $\Delta\mathbf{X}_{\mathbf{DN}}$ in (1)–(3), so that an optimization problem for the design of the FVCs can be formulated using only linear matrix inequality (LMI) constraints. Accordingly, the sizes of $\mathbf{A}_{\mathbf{FF}}$, $\mathbf{B}_{\mathbf{FF}}$, and $\mathbf{C}_{\mathbf{FF}}$ become the same as those of $\mathbf{A}_{\mathbf{DN}}$, $\mathbf{B}_{\mathbf{DN}}$, and $\mathbf{C}_{\mathbf{DG}}$, respectively. Moreover, in (6), the NR-initiating signal $u(t)$ is used as the common input to the FVCs, enabling preemptive compensation for voltage deviations at all DG terminal buses. Note that $u(t–T_d)$ is used to analyze the effect of a communication time delay $T_d$ on FVC performance in Sections IV and V.

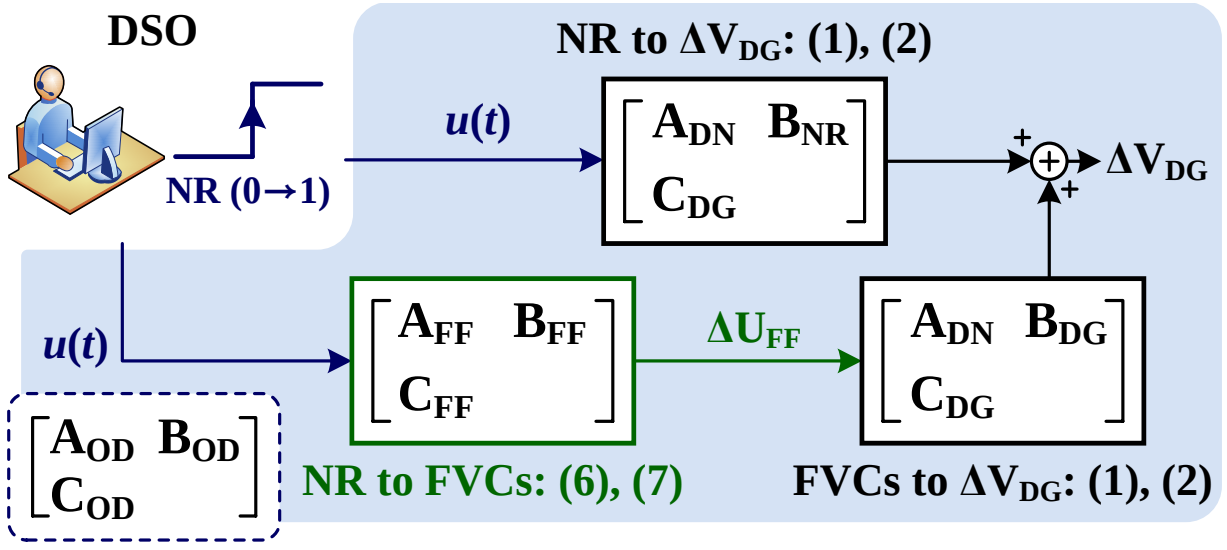

Fig. 2. Dynamic model of a reconfigurable DN with the optimal robust FVCs that are integrated with the existing feedback controllers of DGs.

From (6) and (7), the transfer functions of the FVCs for individual SGs and IGs are obtained as:

$$\mathbf{G}_{\mathbf{FF}}(s) = \left[H_{SG,1}(s)\ \dots\ H_{SG,G}(s)\ M_{IG,1}(s)\ \dots\ M_{IG,L}(s)\right]^T, \quad (8)$$

$$= \mathbf{C}_{\mathbf{FF}} \cdot (s\mathbf{I} - \mathbf{A}_{\mathbf{FF}})^{-1} \cdot \mathbf{B}_{\mathbf{FF}}. \quad (9)$$

The overall dynamics of the reconfigurable DN including the SGs, IGs, and the corresponding FVCs are obtained by combining (1)–(7), as shown in Fig. 2. This yields the dynamic response of $\Delta\mathbf{V}_{\mathbf{DG}}(t)$ to $u(t)$ in the frequency domain, as:

$$\mathbf{G}(s) = \mathbf{C}_{\mathbf{OD}} \cdot (s\mathbf{I} - \mathbf{A}_{\mathbf{OD}})^{-1} \cdot \mathbf{B}_{\mathbf{OD}}, \quad (10)$$

$$\text{where } \mathbf{A}_{\mathbf{OD}} = \begin{bmatrix} \mathbf{A}_{\mathbf{DN}} & \mathbf{B}_{\mathbf{DG}} \cdot \mathbf{C}_{\mathbf{FF}} \\ \mathbf{O} & \mathbf{A}_{\mathbf{FF}} \end{bmatrix},\ \mathbf{B}_{\mathbf{OD}} = \begin{bmatrix} \mathbf{B}_{\mathbf{NR}} \\ \mathbf{B}_{\mathbf{FF}} \end{bmatrix}, \quad (11)$$

$$\mathbf{C}_{\mathbf{OD}} = \begin{bmatrix} \mathbf{C}_{\mathbf{DG}} & \mathbf{O} \end{bmatrix}. \quad (12)$$

Given (10)–(12), the control parameters (i.e., $\mathbf{A}_{\mathbf{FF}}$, $\mathbf{B}_{\mathbf{FF}}$, and $\mathbf{C}_{\mathbf{FF}}$) of the FVCs are determined to minimize the $H_\infty$ norm of $\Delta\mathbf{V}_{\mathbf{DG}}(t)$ (i.e., $||\mathbf{G}(s)||_\infty$) by solving the optimization problem:

**P₁: Problem for the design of optimal robust FVCs**

$$\underset{\mathcal{J},\,\mathcal{L}_{1\text{–}5},\,\mathcal{U}}{\operatorname{argmin}}\ \mathcal{J} \quad (13)$$

$$\text{subject to } \ \mathcal{C}_1 < 0, \quad (14)$$

$$\mathcal{C}_2 = \begin{bmatrix} \mathcal{L}_2 & \mathcal{L}_1 \\ \mathcal{L}_1 & \mathcal{L}_1 \end{bmatrix} > 0,\ \mathcal{L}_1 > 0,\ \mathcal{L}_2 > 0, \quad (15)$$

$$\mathcal{C}_3 = \begin{bmatrix} \mathcal{L}_2 - \mathcal{L}_1 & \mathcal{L}_5^T \\ \mathcal{L}_5 & \mathcal{U} \end{bmatrix} > 0 \text{ for } tr(\mathcal{U}) < \gamma, \quad (16)$$

where the element-wise expression of $\mathcal{C}_1$ in (14) is shown below. Please see Appendix B for the detailed derivation of $\mathbf{P}_1$. Briefly, in (13), $\mathcal{J}$ represents the upper bound of $||\mathbf{G}(s)||_\infty$, which corresponds to the peak value in the frequency response of $\Delta\mathbf{V}_{\mathbf{DG}}(t)$ to $u(t)$. Thus, $\mathbf{P}_1$ is formulated to achieve robust operation of the optimal FVCs in a reconfigurable DN. The constraints (14) and (15) are required to ensure bus voltage stability in the Lyapunov sense. In other words, the optimal solution (i.e., $\mathcal{J}$, $\mathcal{L}_{1\text{–}5}$, and $\mathcal{U}$) of $\mathbf{P}_1$ is obtained such that all poles of $\mathbf{G}(s)$ are located in the left-hand half plane (LHP). Moreover, (16) specifies the upper bound (i.e., $\gamma$) of the total energy of $\Delta\mathbf{U}_{\mathbf{FF}}$; this prevents the excessive operation of the optimal FVCs and hence the DGs.

As shown in (13)–(17), $\mathbf{P}_1$ is a convex optimization problem with a linear objective function and LMI constraints. Therefore, $\mathbf{P}_1$ can be readily solved in real time using a common, off-the-shelf LMI solver. Given a solution of $\mathbf{P}_1$, the optimal control parameters of the FVCs are determined as:

$$\mathbf{A}_{\mathbf{FF}} = (\mathcal{L}_1\mathcal{L}_2^{-1} - \mathbf{I})^{-1}\mathcal{L}_3\mathcal{L}_2^{-1}, \quad (18)$$

$$\mathbf{B}_{\mathbf{FF}} = (\mathbf{I} - \mathcal{L}_1\mathcal{L}_2^{-1})^{-1}\mathcal{L}_4, \text{ and } \mathbf{C}_{\mathbf{FF}} = -\mathcal{L}_5\mathcal{L}_2^{-1}. \quad (19)$$

### *C. Uncertainties in the Estimates of DG and Load Parameters*

As shown in (17), $\mathbf{P}_1$ is formulated using $\mathbf{A}_{\mathbf{DN}}$, $\mathbf{B}_{\mathbf{DG}}$, and $\mathbf{B}_{\mathbf{NR}}$, which include the parameter estimates of the SGs, IGs, voltage-dependent loads, and distribution lines. Note that $\mathbf{C}_{\mathbf{DG}}$ contains only ones and zeros as elements; thus, $\mathbf{C}_{\mathbf{DG}}$ is not associated with the uncertainty. In practice, uncertainties in parameter estimates compromise the estimation accuracies of bus voltage responses to NR and hence the performances of optimal FVCs. In [24]–[26], sensitivity analyses revealed particularly large effects of the exciter amplifier gains $K_A$ of SGs and the filter inductances $L_f$ of IGs on transient variations in their terminal voltages. The load demand $S_r$ to be restored also affects the extent to which the voltages deviate in both the transient state and the steady state after NR [13], [14]. Thus, in this paper, $\mathbf{P}_1$ is extended to consider uncertainties in the estimates of $K_A$, $L_f$, and $S_r$, further enhancing the robustness of the optimal FVCs and their applicability in real DNs.

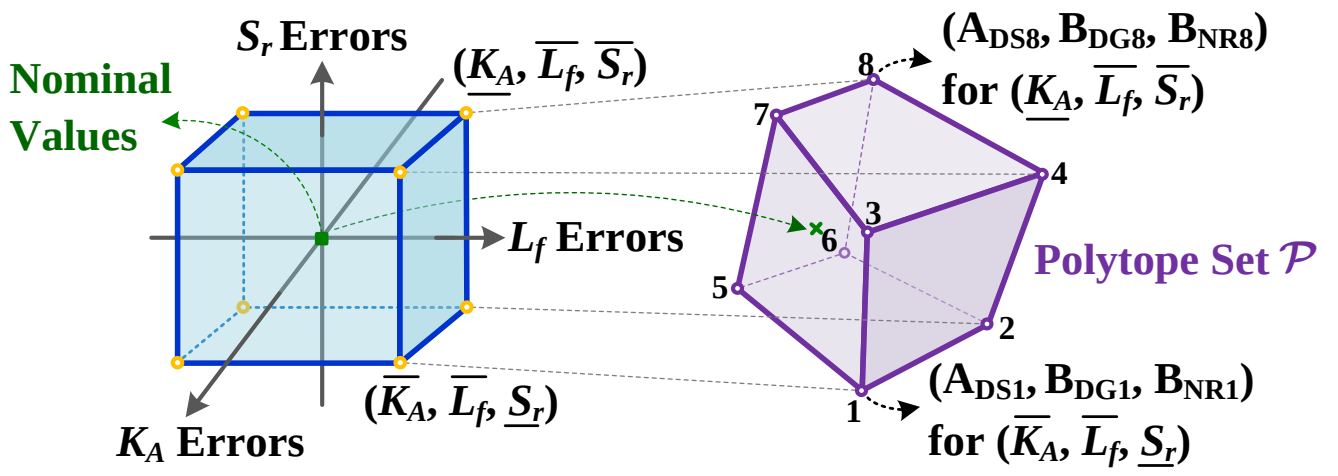

Fig. 3. Polytopic model used to estimate uncertainties in $\mathbf{A}_{\mathbf{DN}}$, $\mathbf{B}_{\mathbf{DG}}$, and $\mathbf{B}_{\mathbf{NR}}$ for the maximum and minimum errors in the estimates of $K_A$, $L_f$, and $S_r$.

Specifically, the effects of uncertainties in the estimates of $K_A$, $L_f$, and $S_r$ on $\mathbf{A}_{\mathbf{DN}}$, $\mathbf{B}_{\mathbf{DG}}$, and $\mathbf{B}_{\mathbf{NR}}$ are first evaluated using the polytopic uncertainty model [27], shown in Fig. 3, where a convex polytope set $\mathcal{P}$ is established as:

$$\mathcal{P} = Co\{[\mathbf{A}_{\mathbf{DN1}}, \mathbf{B}_{\mathbf{DG1}}, \mathbf{B}_{\mathbf{NR1}}], \dots, [\mathbf{A}_{\mathbf{DN}V}, \mathbf{B}_{\mathbf{DG}V}, \mathbf{B}_{\mathbf{NR}V}]\}. \quad (20)$$

In (20), the vertices of $\mathcal{P}$ correspond to $[\mathbf{A}_{\mathbf{DN}v}, \mathbf{B}_{\mathbf{DG}v}, \mathbf{B}_{\mathbf{NR}v}]$ for $v = 1, \dots, V$, determined using the maximum and minimum error percentages in the nominal estimates of $K_A$, $L_f$, and $S_r$. Using the polytopic uncertainty model, the optimal parameters of FVCs are determined to minimize the $H_\infty$ norm of $\Delta\mathbf{V}_{\mathbf{DG}}(t)$ for all inaccurate estimates of $\mathbf{A}_{\mathbf{DN}}$, $\mathbf{B}_{\mathbf{DG}}$, and $\mathbf{B}_{\mathbf{NR}}$ within $\mathcal{P}$, by solving the optimization problem:

**P₂: Extension of P₁ to reflect estimation uncertainty**

$$\underset{\mathcal{J},\,\mathcal{L}_{1\text{–}5},\,\mathcal{U}}{\operatorname{argmin}}\ \mathcal{J} \quad (21)$$

$$\text{subject to } \ \mathcal{C}_{1v} < 0 \ \text{ for } v = 1, \dots, V, \quad (22)$$

$$\text{(15) and (16).}$$

A comparison of (14) and (22) shows that $\mathcal{C}_1$ is extended to $\mathcal{C}_{1v}$ by replacing $\mathbf{A}_{\mathbf{DN}}$, $\mathbf{B}_{\mathbf{DG}}$, and $\mathbf{B}_{\mathbf{NR}}$ in (17) with $\mathbf{A}_{\mathbf{DN}v}$, $\mathbf{B}_{\mathbf{DG}v}$, and $\mathbf{B}_{\mathbf{NR}v}$ in (20), respectively, for all $v$. Extension of $\mathcal{C}_1$ to $\mathcal{C}_{1v}$ further improves the robustness of the optimal FVCs against uncertainties in the estimates of the DG and load parameters. Uncertainties of other parameter estimates also can be readily reflected in the optimization problem $\mathbf{P}_2$ through the polytopic uncertainty modeling. The objective function and the constraints on $\mathcal{C}_{2,\,3}$, $\mathcal{L}_{1,\,2}$, and $\mathcal{U}$ remain the same as in $\mathbf{P}_1$. Thus, $\mathbf{P}_2$ remains a convex optimization problem that can be readily solved. After $\mathbf{P}_2$ is solved, $\mathbf{A}_{\mathbf{FF}}$, $\mathbf{B}_{\mathbf{FF}}$, and $\mathbf{C}_{\mathbf{FF}}$ are determined using (18) and (19), as for $\mathbf{P}_1$.

---

$$\mathcal{C}_1 = \begin{bmatrix} \mathbf{A}_{\mathbf{DN}}\mathcal{L}_2 + \mathcal{L}_2\mathbf{A}_{\mathbf{DN}}^T + \mathbf{B}_{\mathbf{DG}}\mathcal{L}_5 + \mathcal{L}_5^T\mathbf{B}_{\mathbf{DG}}^T & \mathbf{A}_{\mathbf{DN}}\mathcal{L}_1 + \mathcal{L}_2\mathbf{A}_{\mathbf{DN}}^T + \mathcal{L}_5^T\mathbf{B}_{\mathbf{DG}}^T + \mathcal{L}_3^T & \mathbf{B}_{\mathbf{NR}} & \mathcal{L}_2\mathbf{C}_{\mathbf{DG}}^T \\ \mathbf{A}_{\mathbf{DN}}\mathcal{L}_2 + \mathcal{L}_1\mathbf{A}_{\mathbf{DN}}^T + \mathbf{B}_{\mathbf{DG}}\mathcal{L}_5 + \mathcal{L}_3 & \mathbf{A}_{\mathbf{DN}}\mathcal{L}_1 + \mathcal{L}_1\mathbf{A}_{\mathbf{DN}}^T & \mathbf{B}_{\mathbf{NR}} + \mathcal{L}_4 & \mathcal{L}_1\mathbf{C}_{\mathbf{DG}}^T \\ \mathbf{B}_{\mathbf{NR}}^T & \mathbf{B}_{\mathbf{NR}}^T + \mathcal{L}_4^T & -\mathbf{I} & \mathbf{O} \\ \mathbf{C}_{\mathbf{DG}}\mathcal{L}_2 & \mathbf{C}_{\mathbf{DG}}\mathcal{L}_1 & \mathbf{O} & -\mathcal{J}\mathbf{I} \end{bmatrix} < 0 \quad (17)$$

## IV. Small-Signal Analysis

### A. Contribution of the Optimal FVCs to Real-time VR

A small-signal analysis of the proposed VR strategy has been conducted with the optimal FVCs discussed in Section III. In the frequency domain, **G**($s$) [i.e., (10)–(12)] was analyzed for a reconfigurable DN with the model parameters specified in Section V (see Fig. 8 and Table II). Fig. 4 shows that all eigenvalues of **G**($s$) for TSW and SSW operations are placed on the LHP, confirming that the proposed strategy ensures bus voltage stability. Fig. 5 shows the singular value plots (SVPs) of **G**($s$) for the proposed strategy, compared with the SVPs of the conventional strategy using feedback control loops alone. Note that an SVP is an extension of the Bode magnitude plot commonly used to evaluate multi-input and multi-output systems [28]. In Fig. 5, the proposed strategy substantially attenuates bus voltage deviations caused by the SW operations, compared with the conventional strategy, thus improving the real-time VR of a reconfigurable DN.

### B. Sensitivity Analysis

The proposed VR strategy is further analyzed while considering uncertainties in the estimates of $K_A$, $L_f$, and $S_r$, as discussed in Section III-C. For brevity, the SGs and IGs are assumed to exhibit the same error percentages in the nominal values of $K_A$ and $L_f$, respectively; the load units to be restored are assumed to exhibit the same error percentages in $S_r$. Fig. 6 shows the SVPs of **G**($s$) in the proposed and conventional strategies when the error percentages vary by ± 30% [24], [29]. The proposed strategy still leads to lower magnitudes of **G**($s$) and smaller variations in the magnitudes, particularly in the frequency range less than approximately 1.19 × $10^2$ Hz. This verifies the robustness of the proposed strategy against large uncertainties in the estimates of the DG and load parameters.

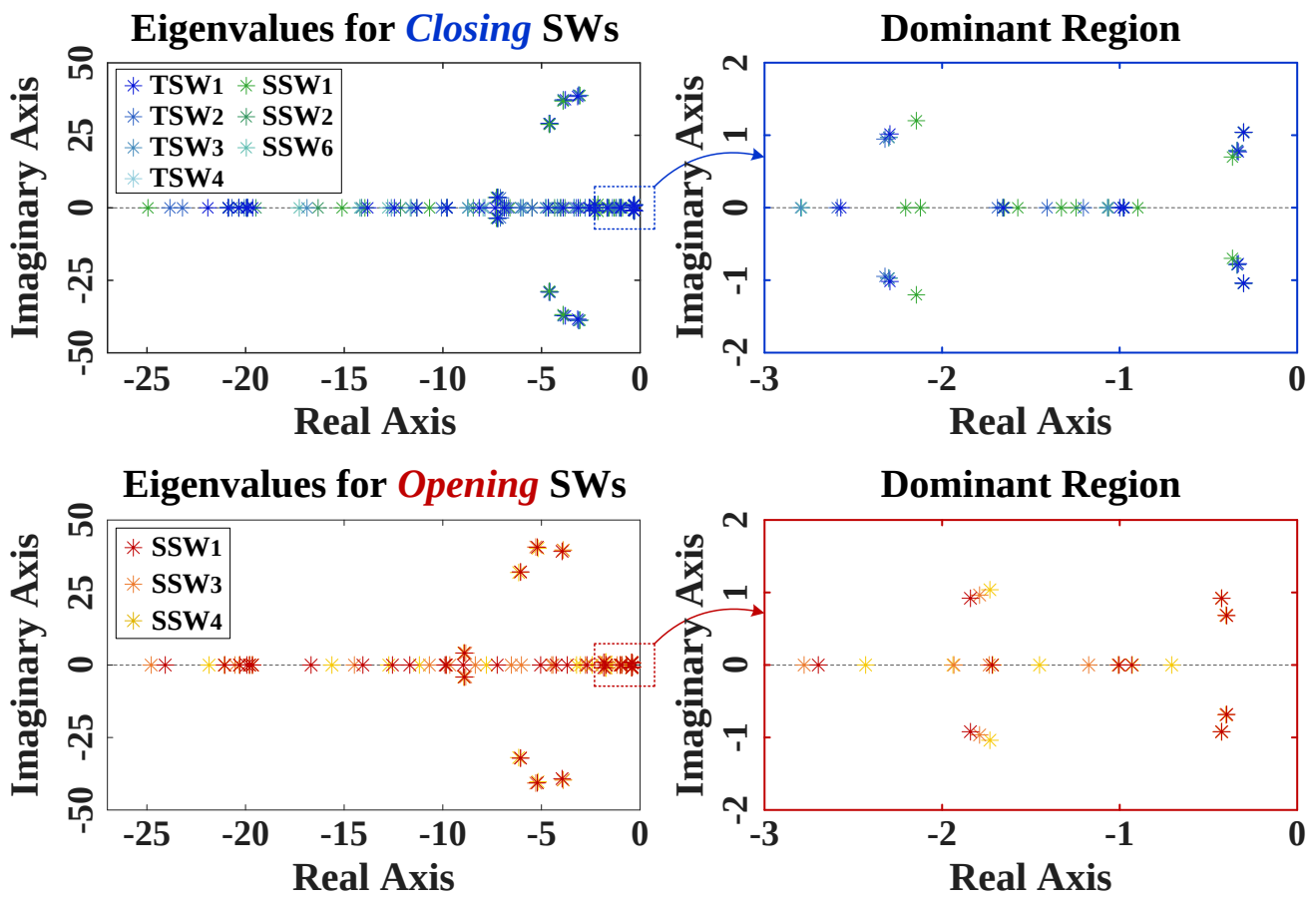

Fig. 4. Eigenvalues of **G**($s$) for the proposed VR strategy.

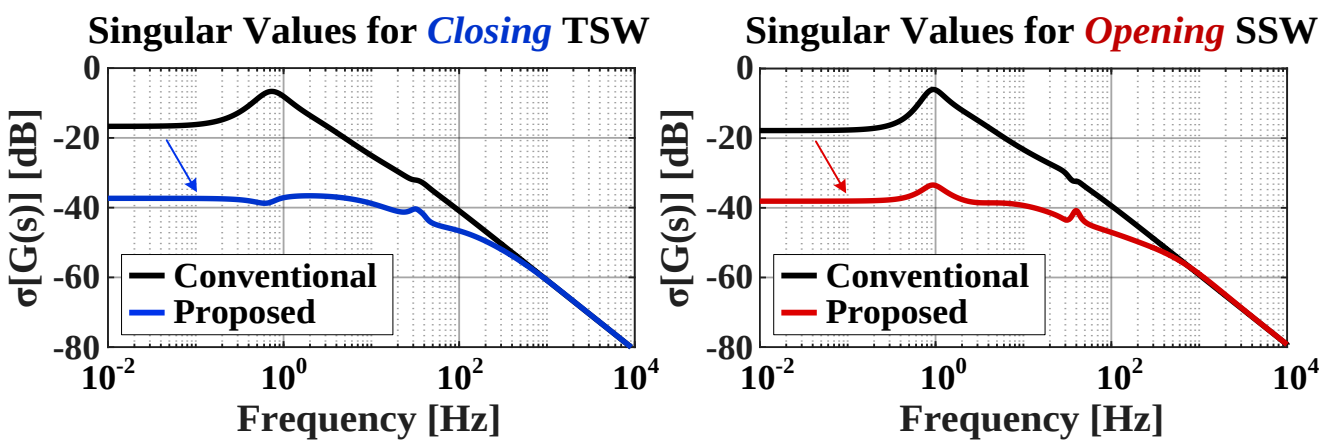

Fig. 5. Singular value plots of the **G**($s$) in the proposed and conventional VR strategies when a TSW and a SSW are closed and opened, respectively.

Sensitivity analysis is also performed when the optimal FVCs respond to NR-initiating signals with a time delay of $T_d$. For $T_d$, the overall dynamics of a reconfigurable DN with the optimal FVCs are:

$$\begin{bmatrix} \Delta\dot{\mathbf{X}}_{\mathbf{DN}}(t) \\ \Delta\dot{\mathbf{X}}_{\mathbf{FF}}(t) \end{bmatrix} = \mathbf{A}_{\mathbf{OD}} \begin{bmatrix} \Delta\mathbf{X}_{\mathbf{DN}}(t) \\ \Delta\mathbf{X}_{\mathbf{FF}}(t) \end{bmatrix} + \begin{bmatrix} \mathbf{B}_{\mathbf{NR}} \\ \mathbf{O} \end{bmatrix} u(t) + \begin{bmatrix} \mathbf{O} \\ \mathbf{B}_{\mathbf{FF}} \end{bmatrix} u(t-T_d). \quad (23)$$

The response of $\Delta\mathbf{V}_{\mathbf{DG}}$ to $u(t)$ then changes from **G**($s$) to:

$$\mathbf{G_d}(s) \approx \mathbf{C_{OD}}\cdot(s\mathbf{I}-\mathbf{A_{OD}})^{-1}\cdot\mathbf{d}(s), \quad (24)$$

$$\text{where } \mathbf{d}(s) = \begin{bmatrix} \mathbf{B}_{\mathbf{NR}} \\ \mathbf{O} \end{bmatrix} + \begin{bmatrix} \mathbf{O} \\ \mathbf{B}_{\mathbf{FF}} \end{bmatrix} \cdot \left( \frac{T_d^2 s^2 - 6T_d s + 12}{T_d^2 s^2 + 6T_d s + 12} \right). \quad (25)$$

Note that the second-order padé approximation of $e^{-sT_d}$ was adopted in (24). Fig. 7 and Table I compare the performances of the proposed and conventional strategies when $T_d$ increases from 0.1 to 0.6 s. Delayed FVC activations render $\Delta\mathbf{V}_{\mathbf{DG}}$ less attenuated, particularly in the frequency range from approximately 1.03 × $10^{-2}$ Hz to 3.29 × $10^2$ Hz, compared with synchronous activation [i.e., **G**($s$)]. However, the proposed strategy still leads to smaller $\|\mathbf{G_d}(s)\|_\infty$ and $\|\mathbf{G_d}(s)\|_2$ values, compared with the conventional strategy, for all $T_d$. In real DNs, the communication time delays were reported to be less than 0.540 s [30], confirming the practical applicability of the proposed strategy. Although it affects the transient voltage responses, $T_d$ has no effect on voltage stability in the proposed strategy because, in (24) and (25), $\mathbf{d}(s)$ is stable in the bounded-input and bounded-output sense. Moreover, the eigenvalues of $(s\mathbf{I} - \mathbf{A}_{\mathbf{OD}})^{-1}$ are the eigenvalues of $\mathbf{G}(s)$, all of which are on the LHP (see Fig. 4).

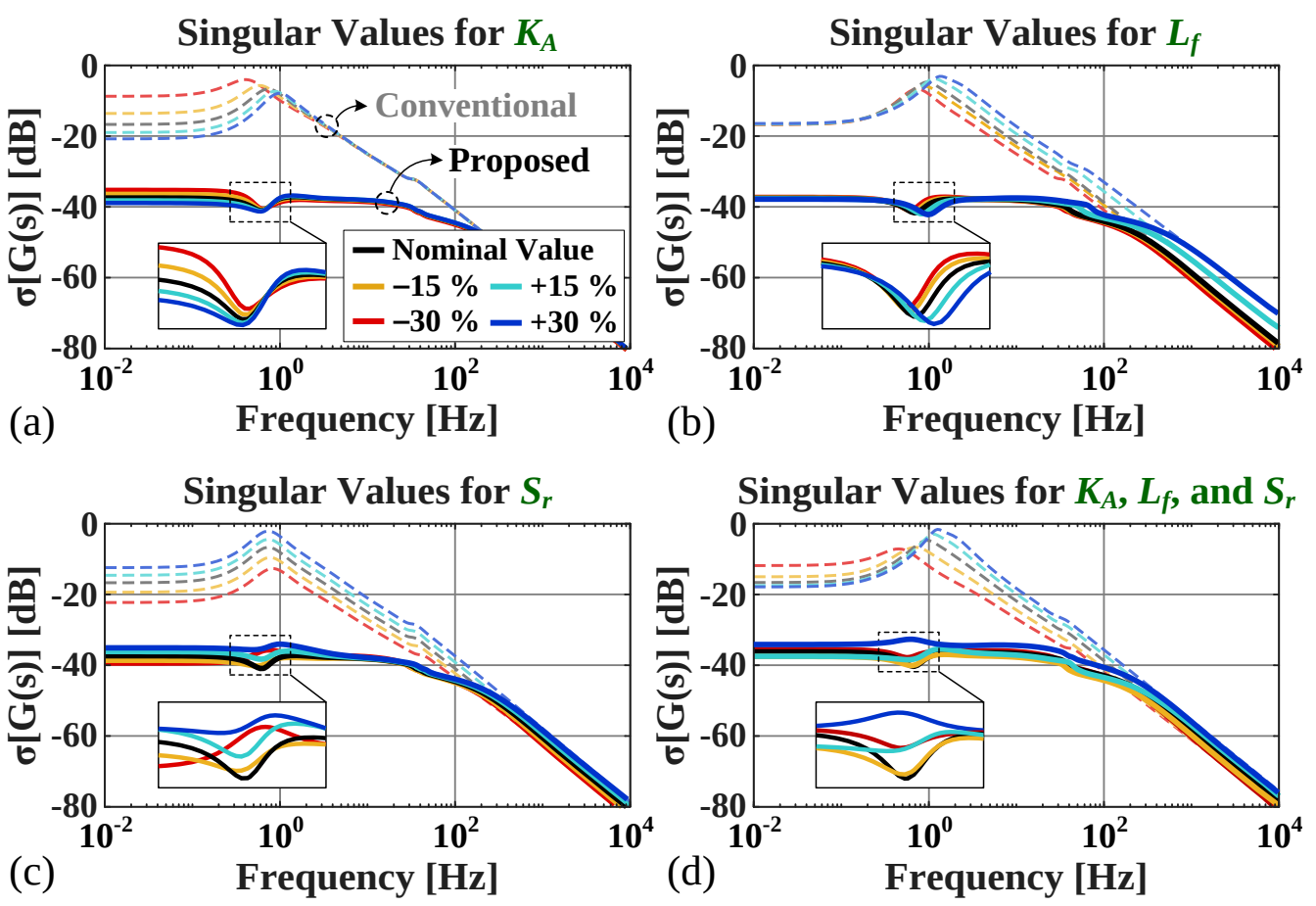

Fig. 6. Singular value plots of **G**($s$) for the proposed and conventional strategies with errors in the estimates of (a) $K_A$, (b) $L_f$, (c) $S_r$, and (d) $K_A$, $L_f$, and $S_r$.

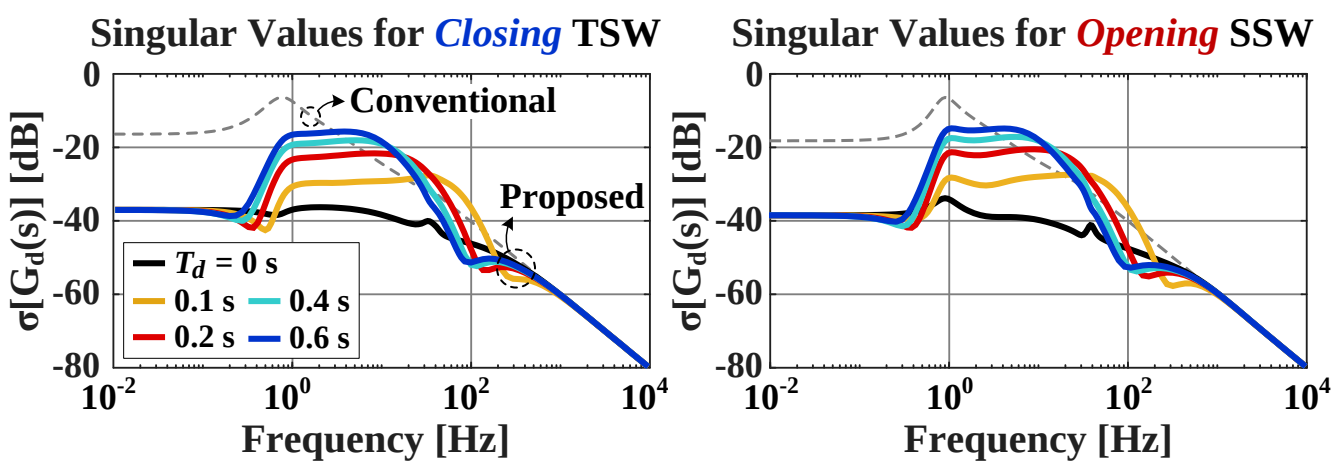

Fig. 7. Singular value plots of $\mathbf{G_d}(s)$ for a communication time delay $T_d$.

Table I. Comparison Between the Proposed and Conventional Strategies for Communication Time Delays

| Comparisons | Conventional | Proposed $T_d$ = 0.1 s | Proposed 0.2 s | Proposed 0.4 s | Proposed 0.6 s |
|---|---|---|---|---|---|
| **Closing a TSW** | | | | | |
| $\|\mathbf{G_d}(s)\|_\infty$ | 0.464 | 0.038 | 0.079 | 0.121 | 0.159 |
| $\|\mathbf{G_d}(s)\|_2$ | 0.332 | 0.175 | 0.234 | 0.268 | 0.295 |
| **Opening a SSW** | | | | | |
| $\|\mathbf{G_d}(s)\|_\infty$ | 0.508 | 0.044 | 0.098 | 0.145 | 0.190 |
| $\|\mathbf{G_d}(s)\|_2$ | 0.366 | 0.205 | 0.285 | 0.327 | 0.357 |

## V. Case Studies and Simulation Results

### A. Test System and Simulation Conditions

The proposed VR strategy was tested on the DN shown in Fig. 8, modeled using the IEEE 37-node Test Feeder [31] with modifications based on [16] and [32]. Table II lists the corresponding modeling parameters. Specifically, the test DN includes SSWs and TSWs that can adaptively change the DN topology in real time. Fig. 8 shows the initial on-off status when two faults occurred at the feeders between Buses 707 and 720 and Buses 711 and 738. Moreover, the test DN contains three SGs and five IGs, with total power capacities of 1.8 MVA and 1.0 MVA, respectively. The total load demand was 2.6 + $j$1.2 MVA and was distributed to the load units connected to all buses. For simplicity, the load units were assumed to have the same ZIP coefficients of 1.5, –2.3, and 1.8 for active power and of 7.4, –12, and 5.6 for reactive power. Three-phase balanced lines were also adopted with impedances set as the average value over the three phases for each line configuration.

In addition, Fig. 9 and Table III show the self-healing scenario, wherein the SSWs and TSWs operate to restore de-energized loads in Areas 1 and 3. The non-critical loads in Area 2 were disconnected to support bus voltages across the DN, and then re-energized after the load restorations in Areas 1 and 3 were completed. In general, SWs are operated one at a time to prevent excessive voltage and frequency fluctuations in the transient state [13]. In this paper, the time intervals between SW operation were set to 10 s. For each SW operation, the optimal control parameters of the FVCs were determined within 2 s by solving $\mathbf{P_2}$ using the MATLAB toolbox YALMIP.

### B. Performance of the Proposed VR Strategy

The proposed and conventional VR strategies were comparatively analyzed for the operations of TSW$_2$ and SSW$_1$ of the test DN. Table IV lists the main features of the proposed strategy (Cases 1 and 2) and conventional strategy (Cases 3 and 4) strategies. Cases 1 and 3 were compared to examine the effects of the optimal FVCs on real-time VR. Errors in the estimates of $K_A$, $L_f$, and $S_r$ were not reflected in Cases 1 and 3. To allow fair comparison, Cases 2 and 4 evaluated the robustness of the optimal FVCs against errors in the parameter estimates, compared with the existing robust controller discussed in [24].

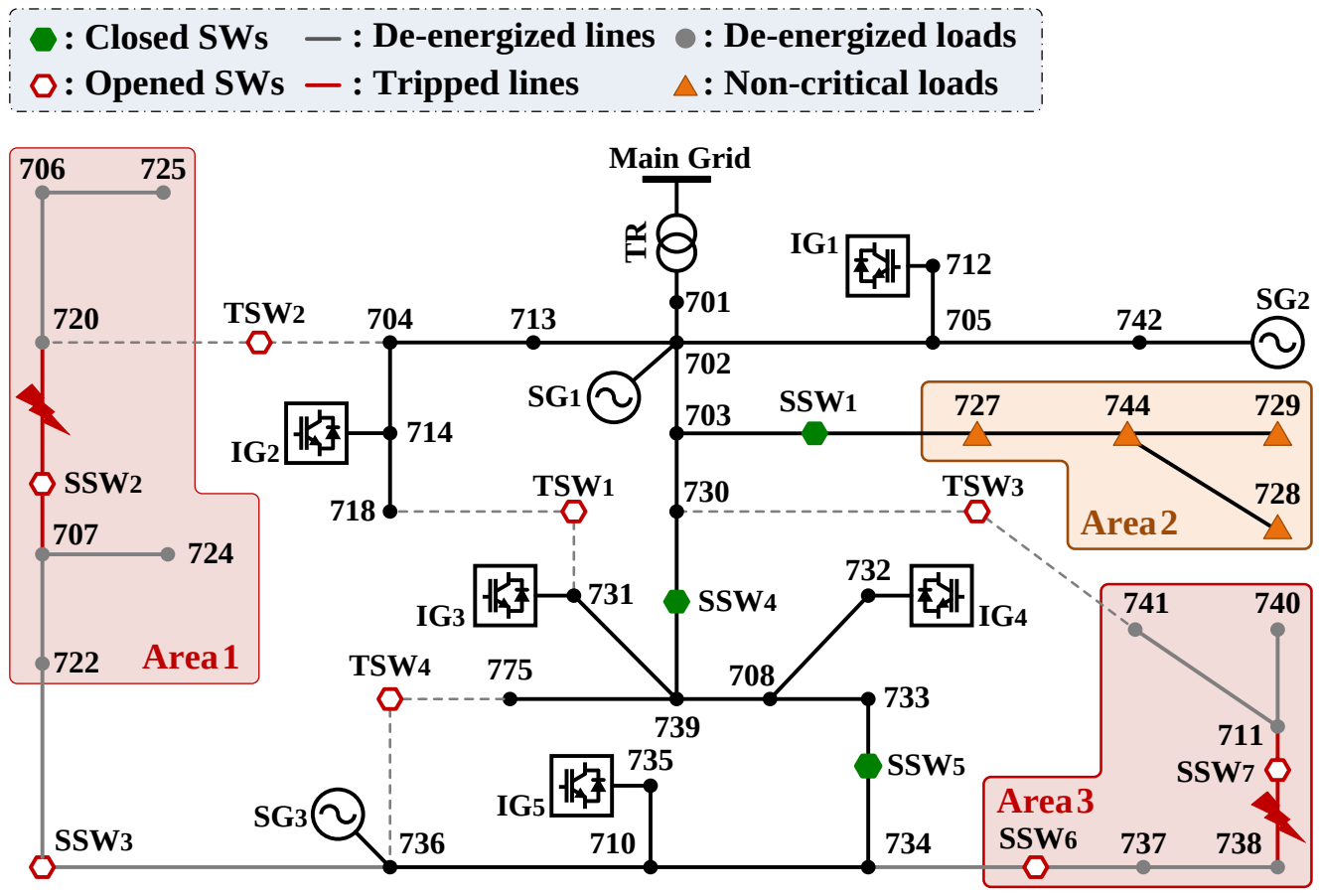

Fig. 8. Single-line diagram of the test DN.

Table II. Network Parameters for the Case Studies

| Device | Description | Parameters | Values |
|---|---|---|---|
| SG units | nominal size and voltage | $S_n$ [MVA], $V_n$ [kV] | 0.6, 2.4 |
| | inertia and damping | $M$ [s], $D$ | 0.5, 0.1 |
| | stator reactances in the $d$ axis | $X_d$, $X'_d$, $X''_d$ [pu] | 2.24, 0.17, 0.12 |
| | stator reactances in the $q$ axis | $X_q$, $X'_q$, $X''_q$ [pu] | 1.1, 0.2, 0.1, 0.04 |
| | open-circuit time constants | $T'_{qo}$, $T''_{qo}$, $T'_{do}$, $T''_{do}$ [s] | 4.5, 0.1, 0.9, 0.03 |
| | field exciter time constants | $T_A$, $T_B$, $T_C$, $T_R$ [s] | 0.02, 5, 1, 0.05 |
| | voltage PI-controller gains | $P_V^{SG}$, $I_V^{SG}$ | 2, 4 |
| | voltage amplifier gain | $K_A$ | 200 |
| IG units | nominal size and DC voltage | $S_n$ [MVA], $V_{DC}$ [V] | 0.2, 380 |
| | filter inductance/resistance | $L_f$ [H], $R_f$ [Ω] | 0.08, 0.91 |
| | transducer time constants | $T_R$ [s] | 0.05 |
| | voltage PI-controller gains | $P_V^{IG}$, $I_V^{IG}$ | 1, 2 |
| | current PI-controller gains | $P_I$, $I_I$ | 20, 30 |
| Loads | rated power demand | $S_L$ [MVA] | 2.6 + $j$1.2 |
| | active power coefficients | $p_Z$, $p_I$, $p_P$ | 1.5, –2.3, 1.8 |
| | reactive power coefficients | $q_Z$, $q_I$, $q_P$ | 7.4, –12, 5.6 |

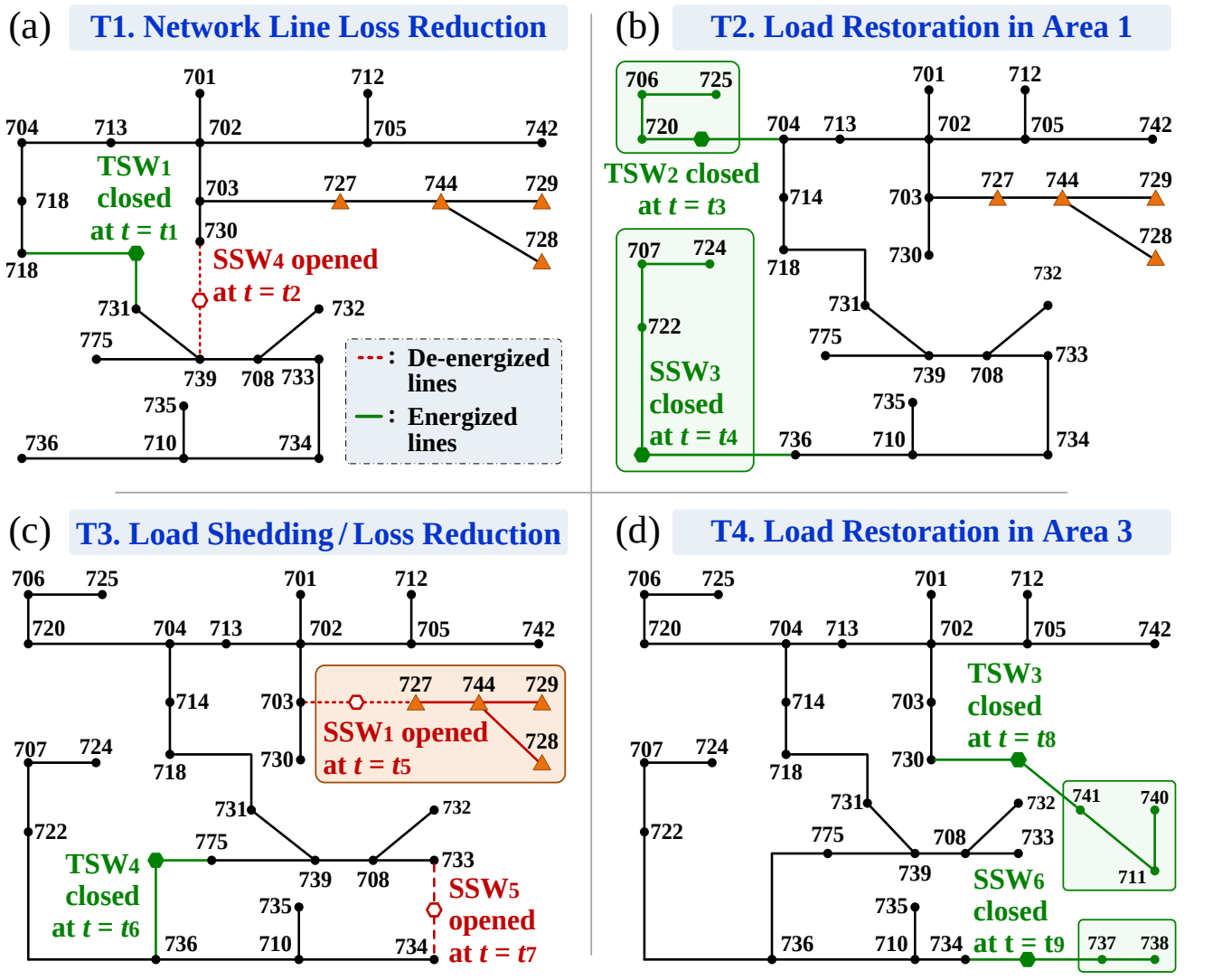

Fig. 9. Variations in the DN topology during the test self-healing scenario: (a) $T_1$, (b) $T_2$, (c) $T_3$, and (d) $T_4$.

Table III. Self-healing Scenario of the Test DN

| Time periods | Operating statuses of the test DN |
|---|---|
| $T_0$ ($t < t_1$) | Faults occur at the feeders between Buses 707 and 720 and between Buses 711 and 738, leading to opening of TSW$_2$ and SSW$_{2,6,7}$ for isolation of the faults. |
| $T_1$ ($t_1 \le t < t_3$) | At $t = t_1$, TSW$_1$ is closed to reduce the line power losses (see Fig. 9(a)). This enables the DGs to secure additional reserve capacity for subsequent load restorations. At $t = t_2$, SSW$_4$ is opened to recover the radial structure of the DN. |
| $T_2$ ($t_3 \le t < t_5$) | TSW$_2$ and SSW$_3$ are closed at $t = t_3$ and $t = t_4$, respectively, to restore the de-energized loads in Area 1 (see Fig. 9(b)). |
| $T_3$ ($t_5 \le t < t_8$) | The non-critical loads in Area 2 are de-energized by opening SSW$_1$ at $t = t_5$ (see Fig. 9(c)) to increase the DG reserve capacity and support the DN voltages. TSW$_4$ and SSW$_5$ then operate at $t = t_6$ and $t = t_7$, respectively, to reduce line power losses, further increasing the reserve capacity. |
| $T_4$ ($t_8 \le t < t_{10}$) | TSW$_3$ and SSW$_6$ are closed at $t = t_8$ and $t = t_9$, respectively, to restore the de-energized loads in Area 3 (see Fig. 9(d)). |
| $T_5$ ($t \ge t_{10}$) | At $t = t_{10}$, SSW$_3$ is closed to restore the non-critical loads in Area 2. The self-healing operation terminates after the faults are investigated and cleared. |

The percentage errors were set to 30% for both Cases 2 and 4, based on the small-signal analysis of Section IV-B.

Fig. 10(a) shows the terminal voltages of $SG_1$, $IG_1$, and $IG_2$ located near $TSW_2$ and $SSW_1$. The proposed strategy significantly reduced voltage deviations caused by the NR-aided load restoration and shedding, compared with the conventional strategies. This led to the considerable reduction of the transient voltage deviations at buses where only loads were connected, as shown in Fig. 10(b). The proposed strategy also decreased the settling times of voltage deviations and hence the time required for consecutive SW operations, facilitating self-healing of the DN. For all buses, $\Delta V_{rms,avg}$, $\Delta V_{pk,max}$, and $\Delta T_{set,max}$ were estimated as:

$$\Delta V_{rms,avg} = \frac{1}{N}\sum_{n=1}^{N}\sqrt{\frac{1}{T}\sum_{t=1}^{T}\Delta V_{n,t}^{2}}\ ,\ \Delta V_{pk,max} = \max\left(\Delta V_{pk,n}\right), \quad (26)$$

$$\text{and } \Delta T_{set,max} = \max\left(\Delta T_{set,n}\right), \quad \text{for } n = 1, \ldots, N. \quad (27)$$

Table V shows that $\Delta V_{pk,max}$, $\Delta V_{rms,avg}$, and $\Delta T_{set,max}$ were smaller for the proposed strategy than for the conventional strategies. The improvement in real-time VR was principally because the proposed FVCs allowed the DGs to respond faster, more robustly, and more accurately to upcoming voltage deviations caused by NR (see Fig. 10(c)), including when the errors in DG and load parameter estimates were large. Fig. 10(d) shows that the active power output profiles of DGs afforded by the proposed and conventional strategies were similar. This implies that the proposed strategy can also be reliably applied to the self-healing of islanded microgrids.

TABLE IV. FEATURES OF THE PROPOSED AND CONVENTIONAL STRATEGIES

| VR strategy | | Description |
|---|---|---|
| Proposed | Case 1 | No uncertainties in the parameter estimates |
| | Case 2 | 30% uncertainties in the parameter estimates |
| Conventional | Case 3 | PI-based output feedback loop [15] |
| | Case 4 | Optimal robust state feedback loop [24] |

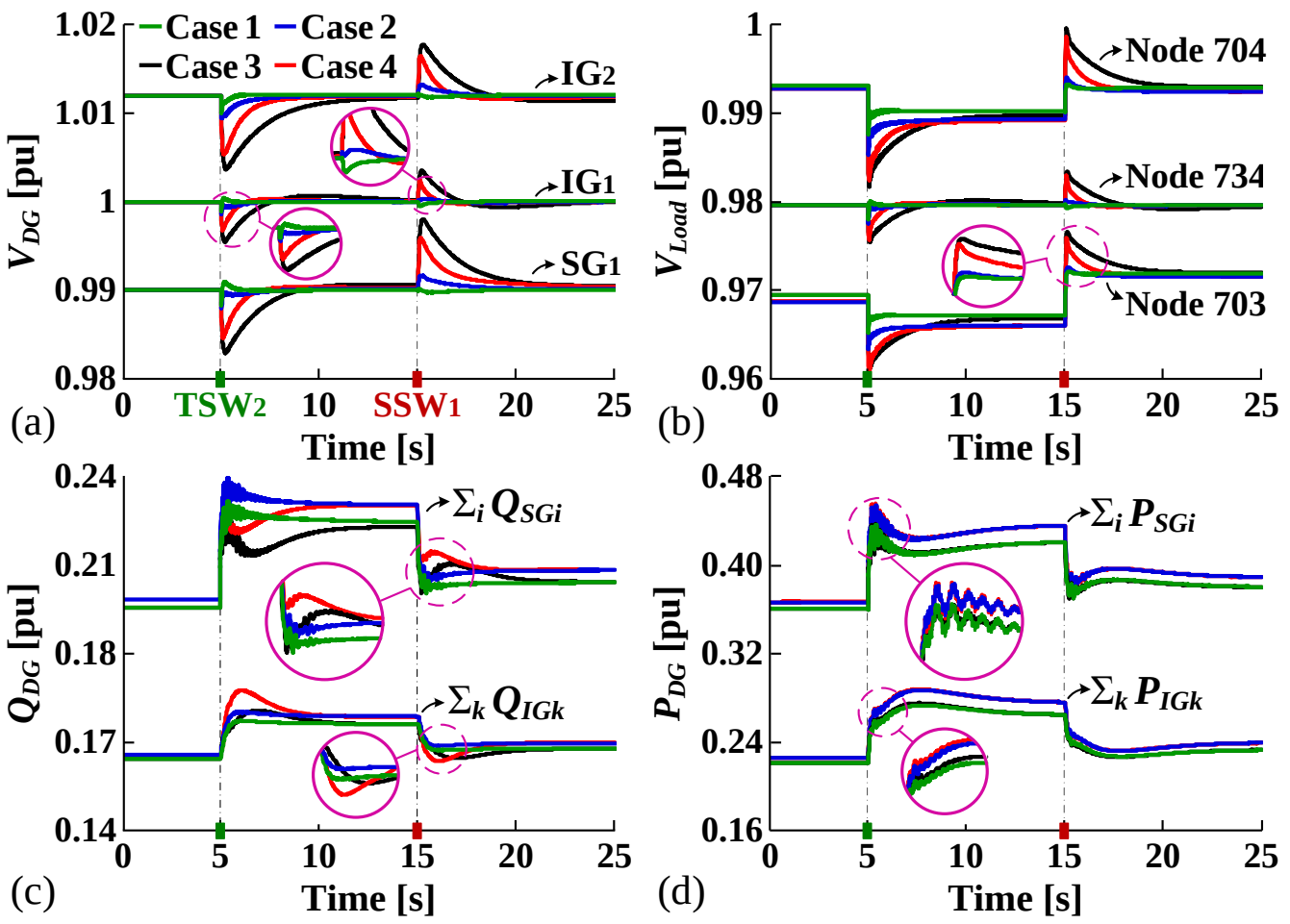

Fig. 10. Comparison of the proposed and conventional VR strategies: (a) $V_{DG}$, (b) $V_{Load}$, (c) $Q_{DG}$, and (d) $P_{DG}$.

TABLE V. COMPARISONS FOR THE STEPWISE LOAD VARIATIONS

| Comparison factors | | Proposed | | Conventional | |
|---|---|---|---|---|---|
| | | Case 1 | Case 2 | Case 3 | Case 4 |
| $\Delta V_{rms,avg}$ | [×10$^{-3}$ pu] | 1.32 | 1.72 | 5.96 | 3.63 |
| $\Delta V_{pk,max}$ | [×10$^{-2}$ pu] | 0.56 | 0.88 | 1.79 | 1.64 |
| $\Delta T_{set,max}$ | [s] | 1.67 | 3.37 | 10.82 | 6.17 |

## *C. Performance in the Self-healing Scenario*

Additional case studies were performed to evaluate the proposed VR strategy with variations over time in the load demand and photovoltaic (PV) power generation [33], [34] (see Fig. 11). The optimal FVCs were developed by reference to the base load demand (i.e., $S_L$ = 2.6 + $j$1.2 MVA). The differences between actual and base load demands were reflected as uncertainties in the network parameter estimates, in addition to uncertainties in the estimates of $K_A$, $L_f$, and $S_r$. Fig. 12 shows the profiles of $V_{DG}$, $V_{Load}$, $Q_{DG}$, and $P_{DG}$ from $T_0$ to $T_5$ in the scenario. In Cases 1 and 2, $\Delta V_{DG}$ and $\Delta V_{Load}$ remain significantly lower at all time periods, compared with Cases 3 and 4, because the optimal FVCs enabled faster and preemptive control of DGs in response to SW operations, thus allowing the existing feedback control loops to better compensate for the remaining $\Delta V_{DG}$. By contrast, in the conventional strategies, DG power outputs were controlled only by the feedback loops; they came into effect after $\Delta V_{DG}$ was already significantly changed by NR. Moreover, Table VI numerically compares the proposed and conventional strategies. For Case 2, $\Delta V_{rms,avg}$ and $\Delta V_{pk,max}$ were 52.3 and 51.9%, respectively, smaller than in Case 4; whereas $\Sigma_i \Delta Q_{SGi,rms}$ and $\Sigma_k \Delta Q_{IGk,rms}$ were only 4.6 and 7.6%, respectively, larger than in Case 4. This implies that the costs incurred by the increased operational stress on SGs and IGs can be adequately compensated by the savings attributable to the improved VR during load restoration.

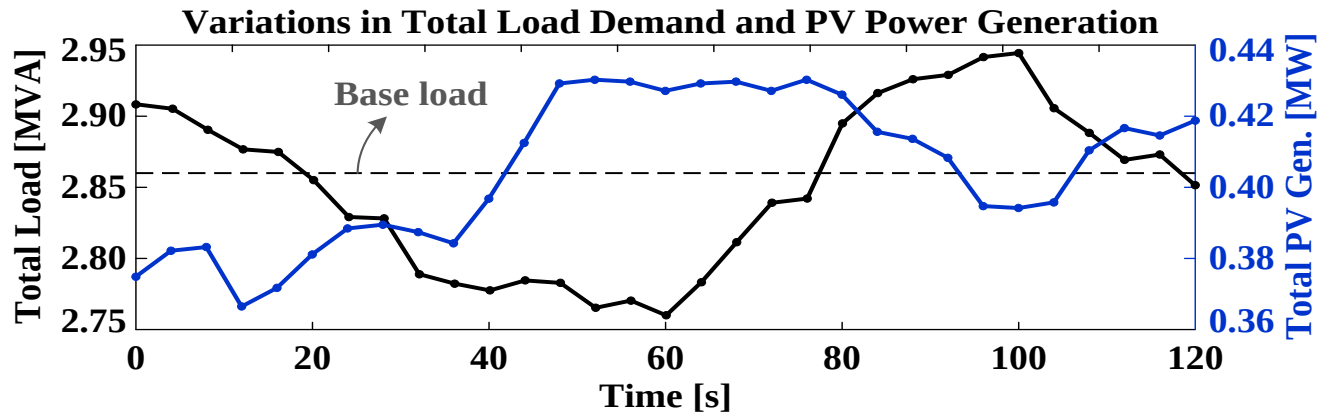

Fig. 11. Continuous variations in the load demand and PV generation.

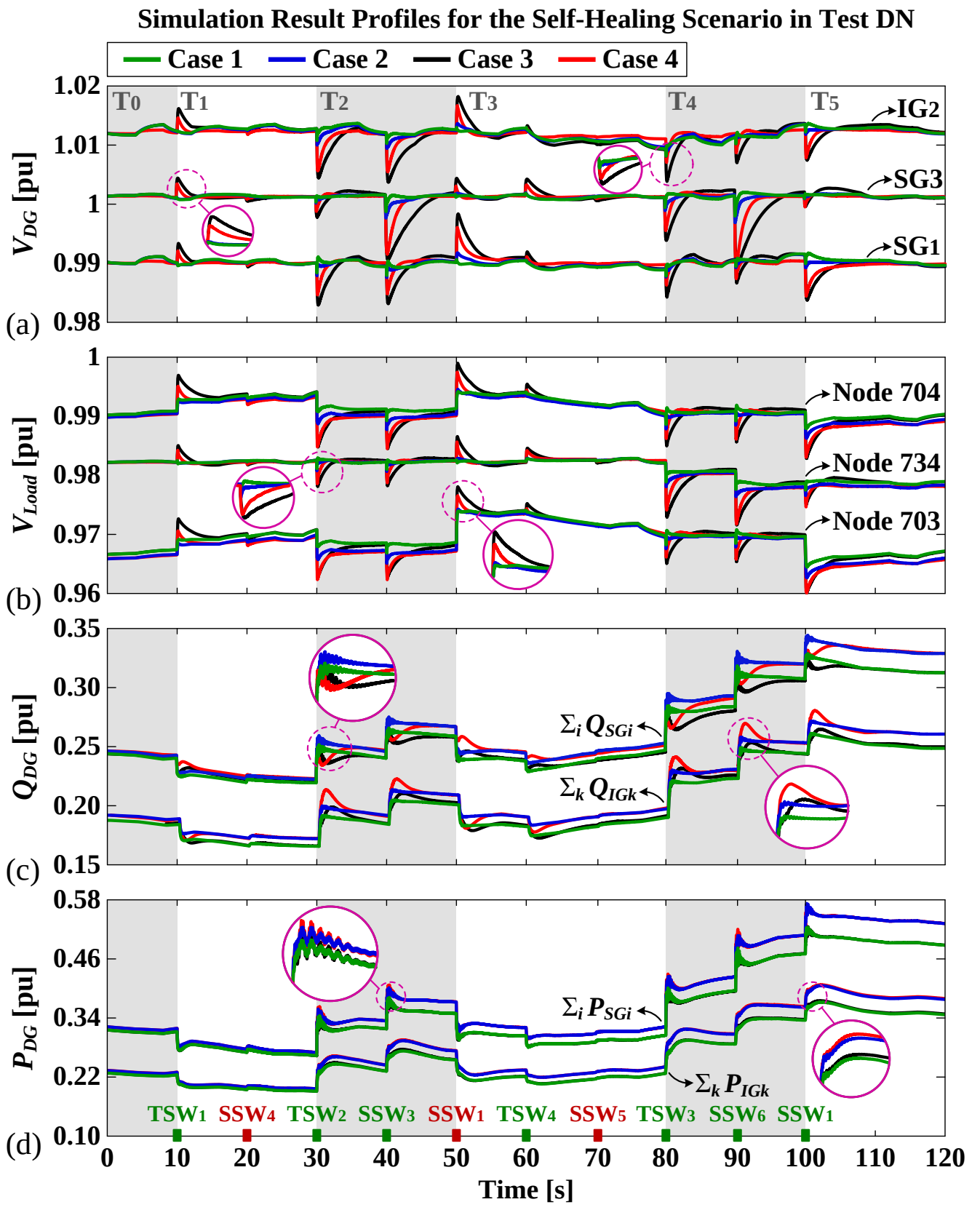

Fig. 12. Comparison of the proposed and conventional VR strategies for the self-healing scenario: (a) $V_{DG}$, (b) $V_{Load}$, (c) $Q_{DG}$, and (d) $P_{DG}$.

TABLE VI. COMPARISONS FOR THE CONTINUOUS LOAD VARIATIONS

| Comparison factors | | Proposed | | Conventional | |
|---|---|---|---|---|---|
| | | Case 1 | Case 2 | Case 3 | Case 4 |
| $\Delta V_{rms,avg}$ | [×10$^{-3}$ pu] | 1.564 | 1.816 | 6.684 | 3.808 |
| $\Delta V_{pk,max}$ | [×10$^{-2}$ pu] | 0.962 | 1.163 | 2.741 | 2.418 |
| $\Sigma_i \Delta Q_{SGi,rms}$ | [pu] | 0.118 | 0.137 | 0.111 | 0.131 |
| $\Sigma_k \Delta Q_{IGk,rms}$ | [pu] | 0.082 | 0.099 | 0.075 | 0.092 |

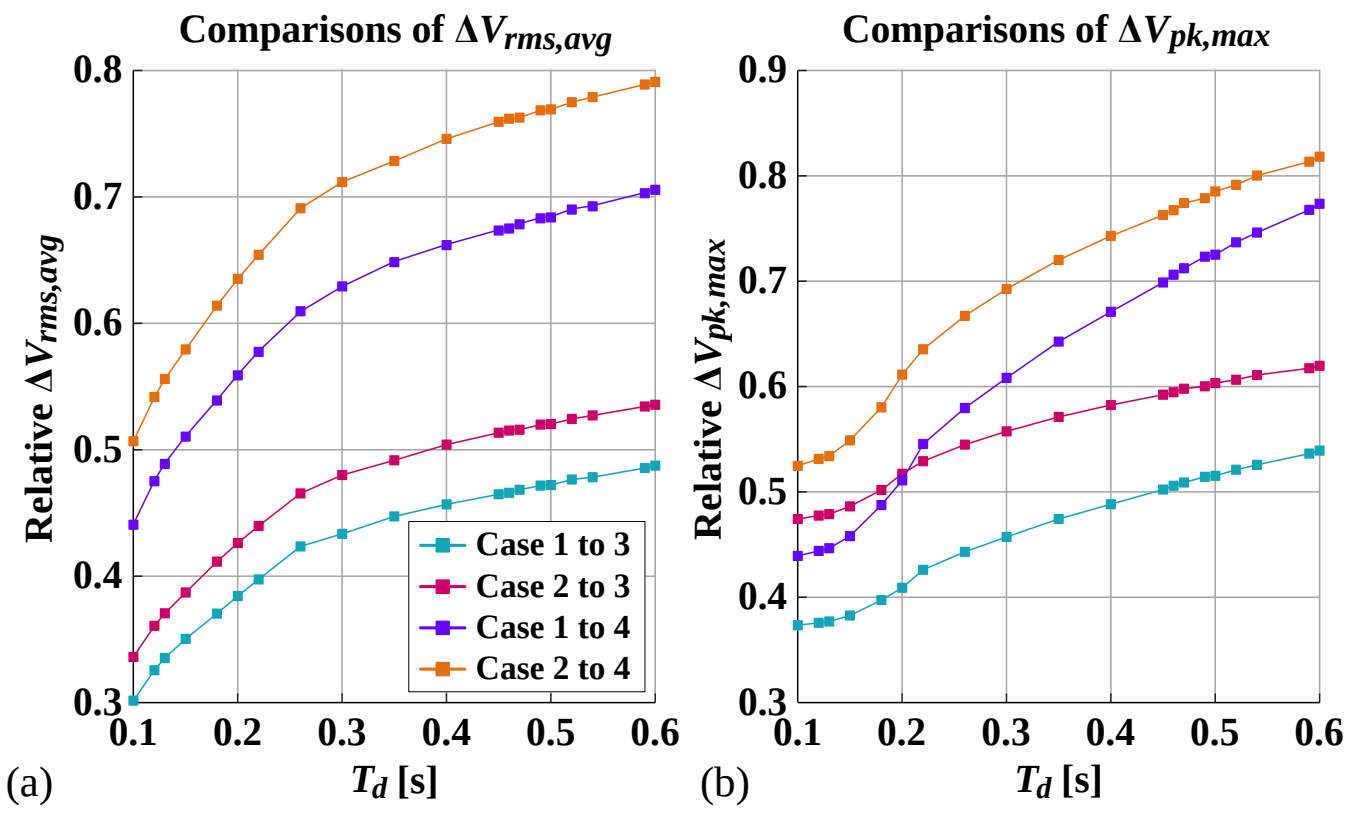

Fig. 13. Relative magnitudes of (a) $\Delta V_{rms,avg}$ and (b) $\Delta V_{pk,max}$ for different communication time delays.

### *D. Sensitivity Analysis for a Communication Time Delay*

The case studies discussed in Section V-C were repeated to analyze the sensitivity of the proposed VR strategy with respect to $T_d$. Uncertainties in the estimates of $K_A$, $L_f$, $S_r$, and $S_L$ were maintained for Cases 2 and 4. Fig. 13(a) shows the ratios of $\Delta V_{rms,avg}$ of the proposed strategy to those of the conventional strategies, when $T_d$ ranged from 0.1 to 0.6 s, as discussed in Section IV-B. Similarly, Fig. 13(b) shows the ratios of $\Delta V_{pk,max}$ values of the proposed and conventional strategies with respect to $T_d$. For all $T_d$, the ratios of both $\Delta V_{rms,avg}$ and $\Delta V_{pk,max}$ remain smaller than 1.0, confirming that the proposed strategy more effectively and robustly reduces bus voltage deviations. This is also consistent with the small-signal analysis results of Fig. 7.

## VI. Conclusions

This paper proposed a new VR strategy for a reconfigurable DN wherein optimal robust FVCs enable SGs and IGs to respond faster and preemptively to real-time voltage deviations caused by NR-aided load restoration. Real-time voltage deviations at DG terminal buses were estimated using a dynamic analytical model of a reconfigurable DN, and then integrated into a robust optimization problem when designing the optimal FVCs. The problem was formulated to minimize voltage deviations with respect to the $H_\infty$ norm while considering uncertainties in the estimates of the DG and load parameters. The results of small-signal analysis confirmed the effectiveness and robustness of the proposed strategy in terms of attenuating low-frequency components of bus voltage deviations. The case studies also revealed that the proposed strategy more effectively reduced the rms and peak-to-peak variations in bus voltages under various grid conditions, compared with conventional strategies using the PI-based feedback controller and the robust feedback controller.

## Appendix

### *A. Modeling a Reconfigurable Network*

The relationship between the real-time *dq*-axis bus voltages and injection currents in the steady-state is:

$$\mathbf{I_0} = \mathbf{Y_B} \cdot \mathbf{V_0}. \tag{A1}$$

After NR is initiated, (A1) changes to:

$$\mathbf{I_0} + \mathbf{\Delta I}(t) = \mathbf{Y_A} \cdot (\mathbf{V_0} + \mathbf{\Delta V}(t)), \tag{A2}$$

where $\mathbf{\Delta V}(t)$ and $\mathbf{\Delta I}(t)$ are the variations in the *dq*-axis voltages and currents, respectively, in the transient state. From (A1) and (A2), $\mathbf{\Delta I}(t)$ can be represented as:

$$\mathbf{\Delta I}(t) = \mathbf{Y_A} \cdot \mathbf{\Delta V}(t) + \mathbf{\Delta I_T}(t), \tag{A3}$$

where $\mathbf{\Delta I_T}(t) = \mathbf{\Delta Y}(t) \cdot \mathbf{V_0}$ and $\mathbf{\Delta Y}(t) = (\mathbf{Y_A} - \mathbf{Y_B}) \cdot u(t)$. Note that NR is considered to be a discrete variation in the admittance matrix (i.e., from $\mathbf{Y_B}$ to $\mathbf{Y_A}$), leading to the step variation $\mathbf{\Delta I_T}(t)$ that arises immediately after the DN topology changes.

In addition, considering the outputs of the FVCs, the dynamics of the SGs and IGs can be represented in aggregated form as:

$$\mathbf{\Delta \dot{X}_{DN}}(t) = \mathbf{A_X} \cdot \mathbf{\Delta X_{DN}}(t) + \mathbf{B_V} \cdot \mathbf{\Delta V}(t) + \mathbf{B_{DG}} \cdot \mathbf{\Delta U_{FF}}(t), \tag{A4}$$

$$\mathbf{\Delta I_{DG}}(t) = \mathbf{C_X} \cdot \mathbf{\Delta X_{DN}}(t) - \mathbf{D_V} \cdot \mathbf{\Delta V}(t), \tag{A5}$$

where the coefficient matrices are obtained from the linearized expressions for the SG and IG models [23]. Moreover, the voltage-dependent loads can be modeled as:

$$\mathbf{\Delta I_L}(t) = \mathbf{D_L} \cdot \mathbf{\Delta V}(t), \tag{A6}$$

where $\mathbf{D_L}$ is a block diagonal matrix, the elements of which are determined based on the ZIP coefficients [23]. Using $\mathbf{\Delta I}(t) = \mathbf{\Delta I_{DG}}(t) + \mathbf{\Delta I_L}(t)$, a dynamic model of the reconfigurable DN can be established by substituting (A5) and (A6) into (A3), as:

$$\mathbf{\Delta V}(t) = \mathbf{Z} \cdot (\mathbf{C_X} \cdot \mathbf{\Delta X_{DN}}(t) - \mathbf{\Delta I_T}(t)), \tag{A7}$$

where $\mathbf{Z} = (\mathbf{Y_A} + \mathbf{D_V} - \mathbf{D_L})^{-1}$. Using (A4) and (A7), the dynamics of $\mathbf{\Delta X_{DN}}$ can then be expressed in a state-space form as:

$$\mathbf{\Delta \dot{X}_{DN}}(t) = \mathbf{A_{DN}} \cdot \mathbf{\Delta X_{DN}}(t) + \mathbf{B_{DG}} \cdot \mathbf{\Delta U_{FF}}(t) + \mathbf{B_{NR}} \cdot u(t), \tag{A8}$$

where $\mathbf{A_{DN}} = \mathbf{A_X} + \mathbf{B_V} \cdot \mathbf{Z} \cdot \mathbf{C_X}$ and $\mathbf{B_{NR}} = -\mathbf{B_V} \cdot \mathbf{Z} \cdot (\mathbf{Y_A} - \mathbf{Y_B}) \cdot \mathbf{V_0}$. In (A8), $\mathbf{\Delta X_{DN}}$ includes $\Delta V_{SGi}$ and $\Delta V_{IGk}$ as:

$$\mathbf{\Delta V_{DG}}(t) = \mathbf{C_{DG}} \cdot \mathbf{\Delta X_{DN}}(t), \tag{A9}$$

where $\mathbf{C_{DG}}$ contains only zeros and ones.

### *B. Robust Optimization with LMI Constraints*

The existence of an upper bound of $\|\mathbf{G}(s)\|_\infty$ is proved as:

***Lemma 1*** [35]: A positive finite $\mathcal{J}$ for $\|\mathbf{G}(s)\|_\infty < \mathcal{J}$ exists if and only if there exists $\mathbf{Q} > 0$ such that

$$\mathcal{C}_{\mathrm{N}} = \begin{bmatrix} \mathbf{QA_{OD}} + \mathbf{A}_{\mathbf{OD}}^T\mathbf{Q} & \mathbf{QB_{OD}} & \mathbf{C}_{\mathbf{OD}}^T \\ \mathbf{B}_{\mathbf{OD}}^T\mathbf{Q} & -\mathbf{I} & \mathbf{O} \\ \mathbf{C_{OD}} & \mathbf{O} & -\mathcal{J}\mathbf{I} \end{bmatrix} < 0. \tag{B1}$$

Using ***Lemma 1***, the FVCs can be designed by solving

**$\mathbf{P_N}$: Nonconvex optimization problem**

$$\underset{\mathcal{J},\mathbf{A_{FF}},\mathbf{B_{FF}},\mathbf{C_{FF}},\mathbf{Q}}{\arg\min} \quad \mathcal{J} \tag{B2}$$

$$\text{subject to} \quad \mathbf{Q} > 0, \;\; \mathcal{C}_{\mathrm{N}} < 0\,. \tag{B3}$$

The solution of $\mathbf{P_N}$ ensures bus voltage stability, because the Lyapunov condition (i.e., $\mathbf{QA_{OD}} + \mathbf{A}_{\mathbf{OD}}^T\mathbf{Q} < 0$) is guaranteed by $\mathcal{C}_{\mathrm{N}} < 0$. To convert $\mathbf{P_N}$ to a convex problem, the decision variables are replaced by the auxiliary variables $\mathcal{R}, \mathcal{N}, \mathcal{U}, \mathcal{V}$, and $\mathcal{L}_{1\text{–}5}$ for the LMI formulation. Specifically, **Q** and $\mathbf{Q}^{-1}$ are partitioned into block matrices as:

$$\mathbf{Q} = \begin{bmatrix} \mathcal{L}_1^{-1} & \mathcal{U} \\ \mathcal{U}^T & \mathcal{R} \end{bmatrix} \text{ and } \mathbf{Q}^{-1} = \begin{bmatrix} \mathcal{L}_2 & \mathcal{V} \\ \mathcal{V}^T & \mathcal{N} \end{bmatrix}, \tag{B4}$$

where all block matrices are of the size of $\mathbf{A_{DN}}$. This renders the size of $\Delta\mathbf{X_{FF}}$ equal to the size of $\Delta\mathbf{X_{DN}}$. In (B4), $\mathcal{R}$, $\mathcal{N}$, $\mathcal{L}_1$, and $\mathcal{L}_2$ are positive definite matrices; $\mathcal{U}$ and $\mathcal{V}$ are arbitrary nonsingular matrices that satisfy $\mathcal{U}\mathcal{V}^T + \mathcal{L}_1^{-1}\mathcal{L}_2 = \mathbf{I}$. One can then set $\mathcal{U} = \mathcal{L}_1^{-1} - \mathcal{L}_2^{-1}$ and $\mathcal{V} = -\mathcal{L}_2$, yielding the equivalent changes of $\mathbf{A_{FF}}$, $\mathbf{B_{FF}}$, and $\mathbf{C_{FF}}$ as:

$$\begin{bmatrix} \mathbf{A_{FF}} & \mathbf{B_{FF}} \\ \mathbf{C_{FF}} & \mathbf{O} \end{bmatrix} \begin{bmatrix} \mathcal{L}_2 & \mathbf{O} \\ \mathbf{O} & \mathbf{I} \end{bmatrix} = \begin{bmatrix} (\mathcal{L}_1\mathcal{L}_2^{-1} - \mathbf{I})^{-1} & \mathbf{O} \\ \mathbf{O} & -\mathbf{I} \end{bmatrix} \begin{bmatrix} \mathcal{L}_3 & -\mathcal{L}_4 \\ \mathcal{L}_5 & \mathbf{O} \end{bmatrix}. \tag{B5}$$

In (B5), $\mathcal{L}_1\mathcal{L}_2^{-1} - \mathbf{I} = -\mathcal{L}_1\mathcal{U}$ is nonsingular, implying that $\mathbf{A_{FF}}$, $\mathbf{B_{FF}}$, and $\mathbf{C_{FF}}$ can be recovered using $\mathcal{L}_{1\text{–}5}$; see (18) and (19).

Using $\mathcal{L}_1$, $\mathcal{L}_2$, and $\mathcal{V}$, the matrices for the congruence transformation are given as:

$$\widetilde{\mathcal{T}} = diag(\mathcal{T}, \mathbf{I}, \mathbf{I}) \quad \text{where } \mathcal{T} = \begin{bmatrix} \mathcal{L}_2 & \mathcal{L}_1 \\ \mathcal{V}^T & \mathbf{O} \end{bmatrix}. \tag{B6}$$

By applying the transformation to $\mathcal{C}_\text{N}$ and **Q** in (B3), $\mathcal{C}_1$ in (14) and $\mathcal{C}_2$ in (15) can be obtained, respectively, as:

$$\mathcal{C}_1 = \widetilde{\mathcal{T}}^T \mathcal{C}_\text{N} \widetilde{\mathcal{T}} \text{ and } \mathcal{C}_2 = \mathcal{T}^T \mathbf{Q} \mathcal{T}. \tag{B7}$$

Considering the parameter uncertainty, $\mathcal{C}_\text{N}$ is extended to $\mathcal{C}_{\text{N}v}$ for $v = 1, \cdots, V$, and then similarly transformed to $\mathcal{C}_{1v}$ in (22) as:

$$\mathcal{C}_{1v} = \widetilde{\mathcal{T}}^T \mathcal{C}_{\text{N}v} \widetilde{\mathcal{T}} \quad \text{for } v = 1, \ldots, V. \tag{B8}$$

Furthermore, the total energy $\Delta\mathbf{U_{FF}}(t)$ is upper-bounded by $\gamma$ if the following holds [36]:

$$\mathcal{U} > \mathbf{C_{FF}} \mathcal{N} \mathbf{C_{FF}}^T \quad \text{for} \quad tr(\mathcal{U}) < \gamma. \tag{B9}$$

Given $\mathbf{Q}\mathbf{Q}^{-1} = \mathbf{I}$, the relationship between the block matrices in (B4) can be specified as:

$$\mathcal{N} = -\mathcal{U}^{-1}\mathcal{L}_1^{-1}\mathcal{V} = \mathcal{L}_2(\mathcal{L}_2 - \mathcal{L}_1)^{-1}\mathcal{L}_2. \tag{B10}$$

Using (19) and (B10), (B9) is expressed in an LMI form as:

$$\mathcal{U} > \mathcal{L}_5(\mathcal{L}_2 - \mathcal{L}_1)^{-1}\mathcal{L}_5^T \quad \text{for} \quad tr(\mathcal{U}) < \gamma. \tag{B11}$$

The upper bound on the total energy of the control input can then be represented as shown in (16) by applying Schur complements to $\mathcal{U} > \mathcal{L}_5(\mathcal{L}_2 - \mathcal{L}_1)^{-1}\mathcal{L}_5^T$ in (B11).